\documentclass[12pt]{article}
\usepackage{epsfig}
\usepackage{amsfonts}
\usepackage{latexsym}
\usepackage{amsmath}
\usepackage{amssymb}
\usepackage{mathrsfs}
\usepackage{hyperref}
\usepackage{cite}
\usepackage{tikz}
\usepackage{caption}
\usepackage{subcaption}
\usepackage{float}

\numberwithin{equation}{section}

\DeclareFontFamily{OT1}{rsfs}{}
\DeclareFontShape{OT1}{rsfs}{m}{n}{
<-7> rsfs5 <7-10> rsfs7 <10-> rsfs10}{}
\DeclareMathAlphabet{\mycal}{OT1}{rsfs}{m}{n}

\newcommand{\be}{\begin{equation}}
\newcommand{\ee}{\end{equation}}
\newcommand{\bea}{\begin{eqnarray}}
\newcommand{\eea}{\end{eqnarray}}
\newcommand{\half}{\frac{1}{2}}
\def\non{\nonumber}
\def\d{\partial}
\def\k{\kappa}

\def\mR{\mathcal{R}}
\def\o{\omega}
\def\O{\Omega}
\def\sl2r{\mathrm{SL(2,\mathbb{R})}}
\def\tR{\tilde{R}}
\def\tT{\tilde{T}}
\def\tP{\tilde{\Phi}}

\begin{document}

\begin{titlepage}
\unitlength = 1mm
\begin{center}

{ \Large {\textsc{Whirling orbits around twirling black holes\\ \vskip.25cm from conformal symmetry }}}

\vspace{1cm}
Shahar Hadar$^\S$ and Achilleas P. Porfyriadis$^\dag$

\vspace{1cm}

{\it $^\S$ Department of Applied Mathematics and Theoretical Physics, \\
 University of Cambridge, Wilberforce Road, Cambridge CB3 0WA, UK}

\vspace{0.5cm}

{\it $^\dag$ Department of Physics, UCSB, Santa Barbara, CA 93106, USA}

\vspace{1cm}

\begin{abstract}
Dynamics in the throat of rapidly rotating Kerr black holes is governed by an emergent near-horizon conformal symmetry. The throat contains unstable circular orbits at radii extending from the ISCO down to the light ring. We show that they are related by conformal transformations to physical plunges and osculating trajectories. These orbits have angular momentum arbitrarily higher than that of ISCO. Using the conformal symmetry we compute analytically the radiation produced by the physical orbits. We also present a simple formula for the full self-force on such trajectories in terms of the self-force on circular orbits.
\end{abstract}

\vspace{1.0cm}

\end{center}
\end{titlepage}

\pagestyle{plain}
\setcounter{page}{1}
\newcounter{bean}
\baselineskip18pt

%%%%%%%%%%%%%%%%%%%%%%%%%%%%%%%%%%%%%%%%%%%%%%%%%%%%%%%%%%%%%%%%%%%%%

\setcounter{tocdepth}{2}

%\today

\tableofcontents

\section{Introduction}
Recently \cite{Abbott:2016blz} gravitational waves from a binary black hole (BH) coalescence were directly observed for the first time by LIGO \cite{LIGO}. This remarkable achievement is an important milestone in gravitational physics. It marks the beginning of a period in which progressively accurate signatures of increasingly numerous strong-gravity events from across the universe will be collected and analysed. The LISA mission \cite{eLISA}, expected to be launched in 2028, will allow accurate measurement of gravitational waves from extreme-mass-ratio-inspirals (EMRI) of compact stellar-size objects into supermassive BHs. Specifically, LISA is expected to provide waveforms from the late, strong gravity dominated, stages of the EMRI that will probe the spacetime geometry in the vicinity of the horizon and precision-test BH uniqueness. 

It is believed that a significant fraction of the supermassive BHs residing at the centers of galaxies are rapidly rotating \cite{Reynolds:2013qqa,Brenneman:2013oba}. 
The enhanced symmetries of such near-extremal Kerr BHs were first studied in the context of quantum gravity: understanding of the near-horizon near-extremal limit of Kerr BHs in \cite{hep-th/9905099} motivated the Kerr/CFT conjecture \cite{Guica:2008mu} which asserts that quantum gravity in the near-horizon region of near-extreme Kerr is holographically dual to a $1+1$ dimensional conformal field theory. 
However, these enhanced symmetries govern \emph{any} physical process near such extreme BHs, including those which are potentially observable up in the sky. In the astrophysical context, there is a growing number of applications in which near-extremal near-horizon symmetries are utilized to study realistic processes in the vicinity of such BHs. Applications include gravitational waves \cite{Porfyriadis:2014fja,Hadar:2014dpa,Hadar:2015xpa,Gralla:2015rpa,Gralla:2016qfw}, magnetospheres \cite{Lupsasca:2014pfa,Lupsasca:2014hua,Compere:2015pja,Gralla:2016jfc}, and, recently, electromagnetic emissions \cite{Porfyriadis:2016gwb}. %It would be very interesting to find more applications for this approach.

In this paper we extend the analyses of gravitational wave signals from near-extremal Kerr EMRIs performed in \cite{Porfyriadis:2014fja,Hadar:2014dpa,Hadar:2015xpa,Gralla:2015rpa}. Refs. \cite{Porfyriadis:2014fja,Hadar:2014dpa,Hadar:2015xpa,Gralla:2015rpa} studied circular and plunging orbits with angular momentum equal to that of the innermost-stable-circular-orbit (ISCO). In this paper we consider orbits with angular momentum arbitrarily higher than that of ISCO.\footnote{Our analysis is carried out for a massless scalar field, but it is readily generalizable to electromagnetism and gravity along the lines of \cite{Hadar:2014dpa}.} We first solve for a 1-parameter family of unstable equatorial circular orbits, labeled by their angular momenta, whose radii extend from the ISCO down to the light ring. We then find a 1-parameter family of conformal mappings that transform them, altogether, to a 2-parameter family of physical orbits, labeled by their energy and angular momentum. These are either plunges or grazing ``zoom-whirl'' orbits. They include plunges that naively (\emph{i.e.} neglecting backreaction) overspin the BH beyond extremality \cite{Jacobson:2009kt,Colleoni:2015afa,Colleoni:2015ena}. For all physical orbits we solve for the corresponding field profile including the observed waveform at future null infinity. Finally, we discuss an application of our results to the study of the self-force, arguing that the self-force on any of the 2-parameter family of physical orbits may also be obtained via the above-mentioned coordinate transformation from the much simpler case of the circular orbit.

The rest of the paper is organized as follows. In section \ref{Trajectories & mapping} we introduce a family of transformations that map unstable circular orbits in the near-horizon geometry of near-extreme Kerr to generic 2-parameter orbits with angular momentum higher than that of ISCO. In section \ref{Radiation} we use these mappings to solve for the field emitted by such orbits, including the observed radiation at future null infinity. In section \ref{Self-force in the plunge} we argue that the full self-force for the generic orbits - not only the radiative part - is given via our mappings from the simpler circular orbit case by a compact analytic formula. The appendix contains details of the derivation of the  conformal mappings employed in this paper.

\section{Trajectories \& mapping} \label{Trajectories & mapping}

In Boyer-Lindquist coordinates, the Kerr metric describing a BH with mass $M$ and angular momentum $J= a M$ is given by ($G=c=1$)
\be
ds^2 = -\frac{\Delta}{\hat{\rho}^2} \left( d\hat{t} - a \sin^2 \theta \, d \hat{\phi} \right)^2 + \frac{\sin^2 \theta}{\hat{\rho}^2} \left( (\hat{r}^2 + a^2) \, d\hat{\phi} - a d\hat{t} \right)^2 + \frac{\hat{\rho}^2}{\Delta} d \hat{r}^2 + \hat{\rho}^2 d \theta^2 \,,
\label{Kerr metric 1}
\ee
where
\be
\Delta = \hat{r}^2 -2 M \hat{r} + a^2 \,, \quad \hat{\rho}^2 =  \hat{r}^2 + a^2 \cos^2 \theta  \,.
\label{Kerr metric 2}
\ee
The horizons are at $r_\pm = M \pm \sqrt{M^2 - a^2}$.
The angular momentum of Kerr BHs is bounded from above by the extremal value $a = M$. Close to extremality the near-horizon dynamics is greatly simplified by the presence of conformal symmetry \cite{hep-th/9905099, Guica:2008mu}. The limit is best explored in Bardeen-Horowitz coordinates
\be
R = \frac{\hat{r} - r_+}{r_+} \,, \quad T = \frac{\hat{t}}{2 M} \,, \quad  \Phi = \hat{\phi} - \frac{\hat{t}}{2 M}  \,.
\label{nearNHEK transformation}
\ee
Parameterizing the deviation from extremality by
\be
\kappa=\sqrt{1-a^2/M^2}\ll 1\,,
\label{near-extremality parameter}
\ee
the near-horizon metric at $R\sim\kappa$ reads
\be
ds^2 = 2M^2 \Gamma(\theta)\left[ -R(R+2\kappa) dT^2 + \frac{dR^2}{R(R+2\kappa)} + d\theta^2 + \Lambda(\theta)^2 (d\Phi + (R+\kappa) dT)^2 \right] \,,
\label{nearNHEK 1}
\ee
where
\be
\Gamma(\theta) = \frac{1+\cos^2\theta}{2} \,, \quad \Lambda(\theta) = \frac{2\sin\theta}{1+\cos^2\theta} \,.
\label{lambda and gamma}
\ee
This metric, which describes the near-horizon geometry of near-extreme Kerr, is referred to as near-NHEK \cite{Bredberg:2009pv} and it solves the Einstein equation on its own. The near-horizon geometry of extreme Kerr, referred to as NHEK, is given by \cite{hep-th/9905099}
\be\label{NHEK}
ds^2 = 2M^2 \Gamma(\theta)\left[ -\tR^2 d\tT^2 + \frac{d\tR^2}{\tR^2} + d\theta^2 + \Lambda(\theta)^2 (d\tP + \tR d\tT)^2 \right] \,.
\ee
This metric solves the Einstein equation on its own as well. Note that NHEK also describes the near-horizon geometry at $\kappa\ll R\ll 1$ in the throat of a near-extreme Kerr (see \emph{e.g.} Appendix A in \cite{Gralla:2015rpa}). NHEK and near-NHEK are different patches 
in the global-NHEK space-time \cite{hep-th/9905099}
\be
ds^2 = 2M^2 \Gamma(\theta)\left[ -(1+\rho^2) d\tau^2 + \frac{d\rho^2}{1+\rho^2} + d\theta^2 + \Lambda(\theta)^2 (d\varphi + \rho d\tau)^2 \right] \,.
\label{globalNHEK}
\ee
The isometry group is $SL(2,\mathbb{R})\times U(1)$ with time translations being part of the $SL(2,\mathbb{R})$.

\begin{figure}[h!]
	\begin{subfigure}[b]{0.32\textwidth}
		\centering
		\resizebox{\linewidth}{!}  
		{
			\begin{tikzpicture}%\chi>0
			\draw[thick] (0,-5) -- (0, 12);
			\draw[thick] (-8,-5) -- (-8, 12);
			
			\filldraw[fill=black!30, thick] (0,-4) to (-4, 0) to (0,4);
			\draw[thick] (-0.1,0) to (0.1,0);
			\draw (0,0) node[right]{$t=0$};
			
			\draw (-2.2,-2.2) node[rotate=-45]{$r=0$};
			\draw (-3.3, 1.2) node[rotate=45]{$r=0$};
			\draw (0.25, -1.75) node[rotate=90]{$r=\infty$};

			\draw[thick] (0,0) to (-5.5, 5.5) to (0,11);
			\draw[thick] (0,0) to [out=135,in=225] (0,11);
			
			\draw (-3.5, 8) node[rotate=45]{$R=0$};
			\draw (-3.5, 3) node[rotate=-45]{$R=0$};
			\draw (0.25, 6) node[rotate=90]{$R=\infty$};
			
			\draw (-2.25, 5.25) node[right]{$R_0$};
			
			\end{tikzpicture}
		}
		\caption{$\chi>0$}
		\label{fig:A}
	\end{subfigure}
	\begin{subfigure}[b]{0.32\textwidth}
		\centering
		\resizebox{\linewidth}{!}  
		{
			\begin{tikzpicture}%-1<\chi<0
			\draw[thick] (0,-5) -- (0, 12);
			\draw[thick] (-8,-5) -- (-8, 12);
			
			\filldraw[fill=black!30, thick] (0,-4) to (-4, 0) to (0,4);
			
			\draw[thick] (-0.1,0) to (0.1,0);
			\draw (0,0) node[right]{$t=0$};
			
			\draw (-2.2,-2.2) node[rotate=-45]{$r=0$};
			\draw (-3.3, 1.2) node[rotate=45]{$r=0$};
			\draw (0.25, -1.75) node[rotate=90]{$r=\infty$};

			\draw[thick] (0,0) to (-2.75, 2.75) to (0,5.5);
			\draw[thick] (0,0) to [out=125,in=235] (0,5.5);
			
			\draw (-1.75, 4.2) node[rotate=45]{$R=0$};
			\draw (-1.4, 1) node[rotate=-45]{$R=0$};
			\draw (0.25, 3.75) node[rotate=90]{$R=\infty$};
			
			\draw (-1, 2.4) node[right]{$R_0$};
			
			\end{tikzpicture}
		}
		\caption{$-1<\chi<0$}
		\label{fig:B}
	\end{subfigure}
	\begin{subfigure}[b]{0.32\textwidth}
		\centering
		\resizebox{\linewidth}{!}  
		{	
			\begin{tikzpicture}%\chi<-1
			\draw[thick] (0,-5) -- (0, 12);
			\draw[thick] (-8,-5) -- (-8, 12);
			
			\filldraw[fill=black!30, thick] (0,-4) to (-4, 0) to (0,4);
			\draw[thick] (-0.1,0) to (0.1,0);
			\draw (0,0) node[right]{$t=0$};
			
			\draw (-2.2,-2.2) node[rotate=-45]{$r=0$};
			\draw (-3.3, 1.2) node[rotate=45]{$r=0$};
			\draw (0.25, -1.75) node[rotate=90]{$r=\infty$};

			\draw[thick] (0,0) to (-1.5, 1.5) to (0,3);
			\draw[thick] (0,0) to [out=120,in=240] (0,3);
			
			\draw (-1, 2.35) node[rotate=45]{$R=0$};
			\draw (-1, 0.65) node[rotate=-45]{$R=0$};
			\draw (0.25, 1.5) node[rotate=90]{$R=\infty$};
			
			\draw (-1.1, 1.5) node[right]{$R_0$};

			\end{tikzpicture}
		}
		\caption{$\chi<-1$}
		\label{fig:C}
	\end{subfigure}
	\caption{Penrose diagrams of the throat geometry for different values of $\chi$. In each case the asymptotically flat region of Kerr is to be attached to the shaded wedge. The wedge bounded by $R=0$ and $R=\infty$ is a near-NHEK patch in coordinates \eqref{nearNHEK 1}. The wedge bounded by $r=0$ and $r=\infty$ is a near-NHEK patch in coordinates \eqref{nearNHEK 2}. The line at $R=R_0$ is a circular orbit in \eqref{nearNHEK 1} which in \eqref{nearNHEK 2} is seen as a plunging orbit in (a) and (b), or as an osculating orbit in (c). }
	\label{figure}
\end{figure}
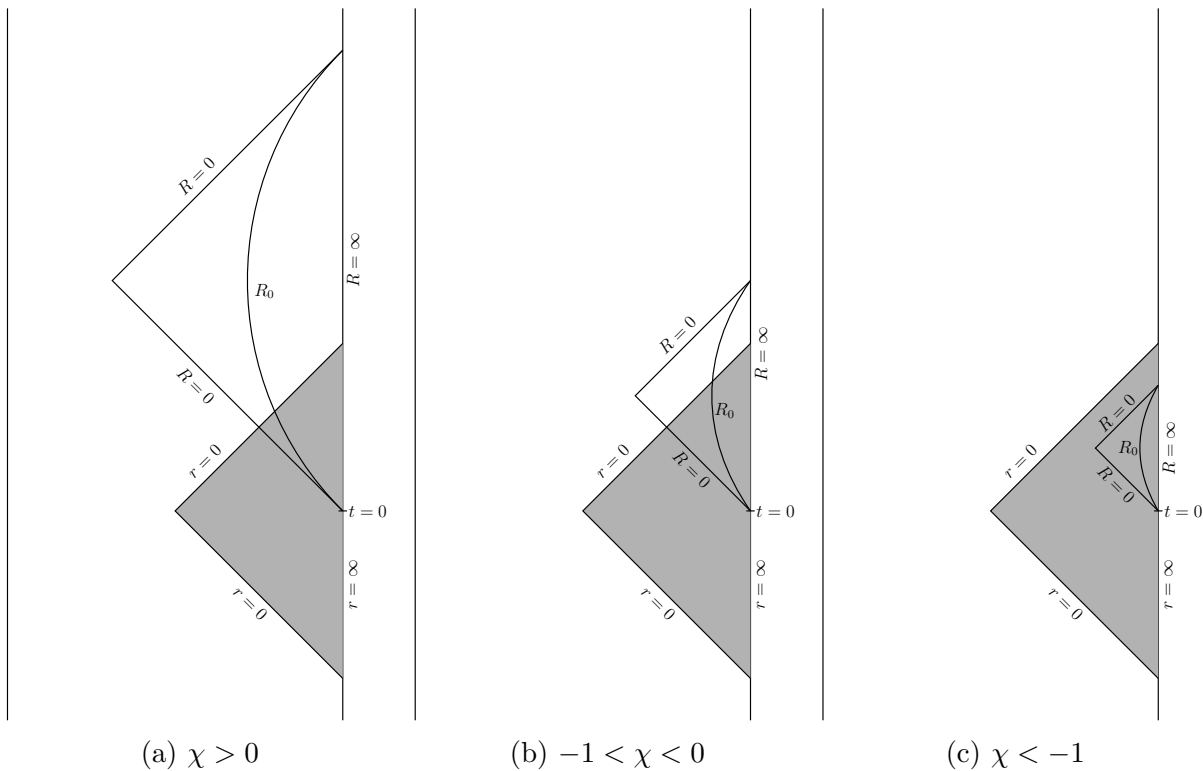

In the near-NHEK geometry \eqref{nearNHEK 1} there exist equatorial circular orbits at any radius above the light ring
\be
R_0 \geq \left(\frac{2}{\sqrt{3}}-1\right) \kappa \,.
\label{r_0}
\ee
Their trajectories are given by
\bea\label{circular orbit}
R &=& R_0 \,,\non \\
\Phi &=& \Phi_0 -\frac{3(R_0+\kappa)}{4} T \,,
\label{circular orbit}
\eea
and they carry near-NHEK energy and angular momentum
\bea
E &=& -\frac{2 M \kappa^2}{\sqrt{3 R_{0}^2 + 6 R_0 \kappa - \kappa^2}} \,,\\  
L &=& \frac{2 M (R_0 + \kappa)}{\sqrt{3 R_{0}^2 + 6 R_0 \kappa - \kappa^2}}\,.
\label{circular e and l}
\eea
These circular orbits are all unstable (see \emph{e.g.} Appendix B in \cite{Hadar:2014dpa}). Therefore, naively, one might think that they are irrelevant for realistic physical situations. This is not the case. Consider the following conformal transformations
\bea \label{transformation}
R &=& \sqrt{r (r + 2 \kappa)} (\sinh \kappa  t+ \chi \cosh \kappa  t)- \chi (r+\kappa)-\kappa \,, \non \\
T &=& \frac{1}{\kappa} \ln \frac{\sqrt{r(r+2\kappa)} \cosh\kappa t - (r+\kappa)}{\sqrt{R(R+2 \kappa)}} \,, \\
\Phi &=& \phi - \frac{1}{2}\ln  \frac{\sqrt{r(r+2\kappa)} - (r+\kappa) \cosh\kappa t + \kappa \sinh\kappa t }{\sqrt{r(r+2\kappa)} - (r+\kappa) \cosh\kappa t - \kappa \sinh\kappa t } \,  \frac{R+2 \kappa}{R} \,, \non
\eea
where $\chi$ is a constant. As explained in the Appendix, these transformations may be thought of as a $\tT\to\tT-\chi$ translation followed by a $\tau\to\tau-\pi/2$ translation. The transformations \eqref{transformation} leave the near-NHEK metric invariant,
\be
ds^2 = 2M^2 \Gamma(\theta)\left[ -r(r+2\kappa) dt^2 + \frac{dr^2}{r(r+2\kappa)} + d\theta^2 + \Lambda(\theta)^2 (d\phi + (r+\kappa) dt)^2 \right] \,,
\label{nearNHEK 2}
\ee
but map the trajectory \eqref{circular orbit} to
\bea \label{general orbit}
t(r) &=& \frac{1}{\kappa} \ln \frac{R_0+\kappa + \chi(r + \kappa) +D }{(\chi+1) \sqrt{r (r + 2 \kappa)}} \,,\non \\
\phi(r) &=& \Phi_0-\frac{3(R_0+\kappa)}{4 \kappa} \ln \frac{r+\kappa +\chi(R_0+\kappa)-D}{(\chi^2-1)\sqrt{R_0(R_0+ 2\kappa)} } + \frac{1}{2} \ln \frac{(R_0+2\kappa)r}{R_0(r+2\kappa)}  \\
&\quad& +\frac{1}{2} \ln \frac{r^2+R_0^2+(\chi-1)R_0r+(\chi+1)^2\kappa^2-(r-R_0)D-\kappa(\chi+1)(D-2r)}{r^2+R_0^2+(\chi-1)R_0r+(\chi+1)^2\kappa^2-(r-R_0)D+\kappa(\chi+1)(D+2R_0)} \,, \non
\eea
where 
\be
D= \sqrt{r^2+2 r (\kappa +\chi (R_0+\kappa))+(\kappa (\chi+1) +R_0)^2}\,.
\ee
These are a 2-parameter family of near-NHEK trajectories which carry near-NHEK energy and angular momentum\footnote{Note that while $l$ coincides with the asymptotic angular momentum $\hat{L}$, the near-NHEK energy $e$ measures the difference $\hat{E}-\hat{L}/(2M)$.}
\bea
e &=&  \frac{2 M \kappa^2 \, \chi}{\sqrt{3 R_{0}^2 + 6 R_0 \kappa - \kappa^2}} \,, \non \\
l &=&  \frac{2 M (R_0 + \kappa)}{\sqrt{3 R_{0}^2 + 6 R_0 \kappa - \kappa^2}}  \,.
\label{plunge e and l}
\eea
For $\chi>-1$ the trajectories \eqref{general orbit} are plunges which start from the near-NHEK boundary at $t=0$ and fall into the future horizon. For $\chi<-1$ they are osculating orbits which penetrate the throat at the near-NHEK boundary at $t=0$ and then exit it at some later finite time. This is illustrated in Figure \ref{figure}.

\section{Radiation field} \label{Radiation}

Consider a point-like scalar-charged object on a geodesic $x_*(\tau)$ which couples to a  massless scalar field $\Psi$. The system is described by the action
\be\label{scalar action}
S = -\frac{1}{2} \int d^4x\sqrt{-g} \left[ \left(\d \Psi\right)^2 + 8 \pi \lambda \Psi \mathcal{S} \right] \,,
\ee
where $\lambda$ is a coupling constant and
\be
\mathcal{S} = -\int d \tau (-g)^{-1/2} \delta^4(x-x_*(\tau)) \,.
\label{scalar source term}
\ee
We are interested in solving for the radiation field produced by an object on the physical orbits \eqref{general orbit}. We may do so analytically by solving first the much simpler case corresponding to the circular orbits \eqref{circular orbit} and then transforming the solution according to \eqref{transformation}.

%\subsection{Circular orbits}\label{Circular orbits}
For the circular orbits \eqref{circular orbit} in the near-NHEK patch \eqref{nearNHEK 1}, decomposing the scalar field as
\be
\Psi = \sum_{\ell,m}  e^{i m (\Phi - \Omega T) } S_{\ell} (\theta) \mathcal{R}_{\ell m} (R) \,,
\label{scalar field decomposition}
\ee
with \[\Omega = -\frac{3(R_0 + \kappa)}{4}\,,\] and substituting into the equations of motion for (\ref{scalar action}) one finds
\be
R(R + 2\k) \mR'' + 2(R+\k) \mR' + \left[ \frac{m^2 (R + \k + \O)^2}{R(R + 2\k)} + m^2 - K \right] \mR = Q R_0 \delta(R-R_0) \,,
\label{radial equation circular orbit}
\ee
\be
\left[\frac{1}{\sin \theta} \d_{\theta} (\sin \theta \d_{\theta}) + K - \frac{m^2}{\sin^2 \theta}  - \frac{m^2}{4} \sin^2 \theta \right] S_{\ell} = 0 \,,
\label{spheroidal wave equation}
\ee
where $Q= -\frac{\sqrt{3} \lambda}{2 M} S_{\ell}(\pi/2)$. The solutions $S(\theta)$ of the (extremal) spheroidal wave equation \eqref{spheroidal wave equation} and their associated eigenvalues $K$ are well known (\emph{e.g.} they are built in Mathematica as \texttt{SpheroidalPS} and \texttt{SpheroidalEigenvalue}).
The solution to (\ref{radial equation circular orbit}) is a Green's function constructed from homogeneous solutions properly matched at $R=R_0$. A particularly useful basis of homogeneous solutions for near-NHEK physics is \cite{Gralla:2015rpa}
\bea
\mathcal{R}_{in} &=& R^{-\frac{i m}{2}( \Omega/\k + 1)} \left( \frac{R}{2\k} + 1 \right)^{\frac{i m}{2}( \Omega/\k - 1)}  {}_2F_1 \left( h-im , 1-h-im ;  1-i m ( \Omega/\k + 1) ;  -\frac{R}{2\k} \right) \,, \non \\
\mathcal{R}_{N} &=&  R^{-h} \left( \frac{2\k}{R} + 1 \right)^{\frac{i m}{2}( \Omega/\k - 1)} {}_2F_1 \left( h-im,h+i m \Omega/\k ; 2h ;  -\frac{2\k}{R} \right) \,,
\label{homogeneous solutions}
\eea
where
\be
h=\frac{1}{2}+ \sqrt{K-2m^2+\frac{1}{4}}\,.
\ee
$\mathcal{R}_{in}$ obeys ingoing boundary conditions at the horizon, $\mathcal{R}_{in}(R\to 0)=R^{-\frac{i m}{2}( \Omega/\k + 1)}$, and $\mathcal{R}_{N}$ obeys Neumann boundary conditions at the near-NHEK boundary, $\mathcal{R}_{N}(R\gg\k)=R^{-h}$.
The solution to equation \eqref{radial equation circular orbit} with these boundary conditions is given by
\be
\mR_{c} = \frac{Q R_0}{A (1-2h)} \mR_{in}(R_{<}) \mR_{N}(R_{>}) \,,
\label{circular solution}
\ee
where $R_{<}$ and $R_{>}$ are the lesser and greater of $R_0$ and $R$, respectively, and
\be
A = \frac{\Gamma(2h-1) \Gamma(1-i m ( \Omega/\k + 1))}{\Gamma(h - i m) \Gamma(h-i m \Omega/\k)} (2\k)^{1-h-i m( \Omega/\k + 1)/2} \,.
\label{A def}
\ee

\subsection{Physical orbits}\label{Generic orbits}

The solution for the physical orbits \eqref{general orbit} in the near-NHEK patch \eqref{nearNHEK 2} is obtained analytically by applying the transformation \eqref{transformation} on the above solution for the circular orbit. 
This yields the near-NHEK solution with causal boundary conditions at the horizon $r=0$ and Neumann at $r\gg\k$.
Indeed, for $r\gg\k$ and fixed $t, \phi$ the transformation \eqref{transformation} reduces to
\bea
R &\approx& r \left[ \sinh \k t + \chi (\cosh\k t-1) \right] \gg\k \,, \non \\
T &\approx& - \frac{1}{\k} \ln\left(\coth\frac{\k t}{2}+\chi\right) \,,  \\
\Phi &\approx& \phi \,, \non
\label{diffeo at boundary}
\eea
so that plugging into \eqref{circular solution} and \eqref{scalar field decomposition} we obtain, for $t>0$,
\be
\Psi(r\gg\k) = \frac{1}{\sqrt{2 \pi}} \int d\o \sum_{\ell m} \frac{Q  R_0  \mathcal{R}_{in}(R_0)}{\k A (1-2h)} \mathcal{N} e^{i (m \phi - \o t)} S_{\ell}(\theta) r^{-h} \,,
\label{boundary radiation 1}
\ee
where
\be
\mathcal{N} = \frac{1}{\sqrt{2\pi}} \int_{0}^{\infty} dy \, e^{i \frac{\o}{\k} y} \left( \coth \frac{y}{2} + \chi \right)^{-\frac{3im(R_0+\k)}{4\k}}\left[ \sinh y + \chi (\cosh y -1) \right]^{-h}  \,.
\label{boundary radiation 2}
\ee

In order to find the radiation field at asymptotically flat infinity one needs to match this near-NHEK solution to a solution in the far Kerr geometry. This may be done using the method of matched asymptotic expansions (MAE). For a detailed presentation of the method applied to a different but similar solution in near-NHEK we refer the reader to section 4.4 in \cite{Hadar:2014dpa}. Here we will only review the general idea and necessary definitions and give the final result for the orbits \eqref{general orbit} considered in this paper.

Consider a scalar field on the full near-extreme Kerr geometry in Boyer-Lindquist coordinates and expand in modes
\be
\Psi = \frac{1}{\sqrt{2 \pi}} \int d \hat{\o} \sum_{\ell m} e^{i(m \hat{\phi}-\hat{\o}\hat{t})} \hat{S}_{\ell}(\theta) \hat{\mathcal{R}}_{\ell m \hat{\o}}(\hat{r}) \,,
\label{Kerr expansion}
\ee
with $\hat{S}_{\ell}(\theta)$ the standard Kerr spheroidal harmonics. Identify the near-NHEK geometry which contains the orbits studied in this paper as the region given by $r=(\hat{r}-r_+)/r_+\ll 1$.
Let the dimensionless Hawking temperature and rescaled near-superradiant frequency be
\be
\tau_H = {r_+-r_-\over r_+}\,,\quad n = 4M\frac{\hat{\o} - m \Omega_H}{\tau_H}\,,
\ee
with $\Omega_H = a/(2 M r_+)$ the horizon angular velocity.\footnote{ Note that to leading order $\tau_H=2\k$.} The solution for the field in the far Kerr region, $r\gg max(\tau_H,n\tau_H)$, with no incoming radiation from past null infinity is given by equation (4.14) in \cite{Hadar:2014dpa}. Identifying $\omega=(n-m)\k$ the near-NHEK solution may be matched to the far Kerr solution in the overlap region $max(\tau_H,n\tau_H)\ll r\ll 1$. It should be noted that matching to the purely outgoing solution at null infinity requires modifying the Neumann boundary condition on the near-NHEK solution to so called ``leaky'' boundary conditions \cite{Porfyriadis:2014fja}. These are such that at the boundary of near-NHEK, which is in the matching region, we have the appropriate ratio of Dirichlet to Neumann modes that ensures the correct amount of radiation leaks through the near-NHEK boundary and reaches future null infinity. In the end, the result for the waveform at future null infinity, for $m>0$, is given by:
\bea\label{waveform at future null infinity}
&&\hat{\mathcal{R}}_{\ell m \hat{\o}}(\hat{r}\to\infty)= 
\frac{2 \sqrt{3} \lambda}{\pi} (-2i\k)^{h-1}  e^{\frac{\pi m}{2}}  m^{h-1+i m}  \left(\frac{R_0}{2\k}\right)^{1 - \frac{i m}{2}( \Omega/\k + 1)}  \left( \frac{R_0}{2\k} + 1 \right)^{\frac{i m}{2}( \Omega/\k - 1)} \times \non \\
&&\quad\times\,  S_{\ell}(\pi/2) \, {}_2F_1 \left( h-im , 1-h-im ;  1-i m ( \Omega/\k + 1) ;  -\frac{R_0}{2\k} \right)  \times \non \\
&&\quad\times\, \frac{\Gamma(1-2h) \Gamma(h - i m  \Omega/\k)}{\Gamma(2h-1) \Gamma(1-i m (\Omega/\k + 1))} \Gamma(h-i m)^2  \sin(2 \pi h) \, \mathcal{N} \times \\
&&\quad\times\, \left[ 1-(-2 i m \k)^{2h-1} \frac{\Gamma(1-2h)^2  \Gamma(h-i m)^2}{\Gamma(2h-1)^2  \Gamma(1-h-i m)^2} \frac{\Gamma(h-i(n-m))}{\Gamma(1-h-i(n-m))} \right]^{-1} r^{-1+i m} e^{i m r/2}\,. \non
\eea

\section{Self-force} \label{Self-force in the plunge}

In this section we propose another application of the near-extreme conformal symmetry: computation of the self-force (SF) on generic equatorial orbits in the near-horizon region. We deal with the scalar case, as in throughout this paper, but a generalization to gravity should be possible along lines similar to those taken in \cite{Hadar:2014dpa}. In the gravitational case, reconstruction of the metric is required to find the full SF (see \cite{Merlin:2016boc} for recent progress).

The scalar SF is given by
\bea
F^{(SF)}_a = \nabla_a \Psi^{R} \, \, ,
\label{SF def}
\eea
where $\Psi^{R}$ is the regular piece of the field as defined by Detweiler and Whiting \cite{Detweiler:2002mi}. The transformation (\ref{transformation}) is a diffeomorphism: locally, it is just a change of coordinates. Since the singular-regular decomposition is defined locally and is insensitive to changes in the global spacetime structure, the same decomposition will hold after performing (\ref{transformation}). Now, the quantity $F^{(SF)}_a$ transforms to the plunge coordinates as a proper vector:
\bea
F^{(plunge)}_a = \frac{\d X^A}{\d x^a} F^{(circular)}_A \, \, ,
\label{self-force transfomration}
\eea
where $X^A$ stands for the coordinates in \eqref{nearNHEK 1} and $x^a$ stands for the coordinates in \eqref{nearNHEK 2}. $X^A(x^a)$ is given by (\ref{transformation}). This gives a \emph{remarkably simple formula for the SF} on generic near-horizon orbits from the circular SF.

It is important to note that, to leading order in the deviation from extremality, the SF computation is insensitive to the boundary conditions in the asymptotically flat region\footnote{Apart from certain fine-tuned boundary conditions that eliminate the Neumann mode at the boundary.}. It is possible, therefore, to use the solution (\ref{circular solution}) for SF computations without needing to worry about matching as in \eqref{waveform at future null infinity}. It will be interesting to numerically test our formula (\ref{self-force transfomration}) for the self-force.

\section*{Acknowledgements}
We are grateful to Leor Barack for useful conversations. SH is supported by the Blavatnik Postdoctoral Fellowship. APP is supported by NSF grant PHY-1504541.
\appendix
\section{Conformal mappings of near-NHEK}
In this appendix we give some details in relation to the conformal transformation \eqref{transformation} employed in this paper in order to map unstable circular orbits in near-NHEK to physical plunges and osculating orbits. In particular, we show how \eqref{transformation} may be obtained by composing a NHEK time translation $\tilde{T}\to\tilde{T}-\chi$ with a global time translation $\tau\to\tau-\pi/2$.

\begin{figure}[h!]
	\centering
	
	\resizebox{0.3\textwidth}{!}
	{
	    	
	\begin{tikzpicture}
	\draw[thick] (0,-1) -- (0, 17);
	\draw[thick] (-8,-1) -- (-8, 17);
		
	\draw[thick, dashed] (0,0) to (-8, 8) to (0,16);
		
	\draw[thick] (0,0) to (-4, 4) to (0,8);
		
	\draw[thick] (0,0) to (-5.5, 5.5) to (0,11);
	
	\draw (-1.75,2.25) node[rotate=-45]{$\tilde{r}=0$};
	\draw (-1.75, 5.75) node[rotate=45]{$\tilde{r}=0$};
	\draw (-0.3, 4) node[rotate=90]{$\tilde{r}=\infty$};
	
	\draw (-3.25, 7.25) node[rotate=45]{$R=0$};
	\draw (-4.5, 5) node[rotate=-45]{$R=0$};
	\draw (-0.3, 9.25) node[rotate=90]{$R=\infty$};
	
	\draw (-5.75, 9.75) node[rotate=45]{$\tilde{R}=0$};
	\draw (-6.4, 7) node[rotate=-45]{$\tilde{R}=0$};
	\draw (-0.35, 13.25) node[rotate=90]{$\tilde{R}=\infty$};
	
	\end{tikzpicture}
	
	}

	\caption{Penrose diagram illustrating the setup for deriving the transformation \eqref{NHEK time translation}. The solid wedge bounded by $\tilde{r}=0$ and $\tilde{r}=\infty$ is near-NHEK. The solid wedge bounded by $R=0$ and $R=\infty$ is also near-NHEK. Both are embedded in the dashed NHEK wedge bounded by $\tilde{R}=0$ and $\tilde{R}=\infty$. The near-NHEK wedges are related to each other by a NHEK time translation $\tilde{T}\to\tilde{T}-\chi$.}
	\label{appfig1}
\end{figure}
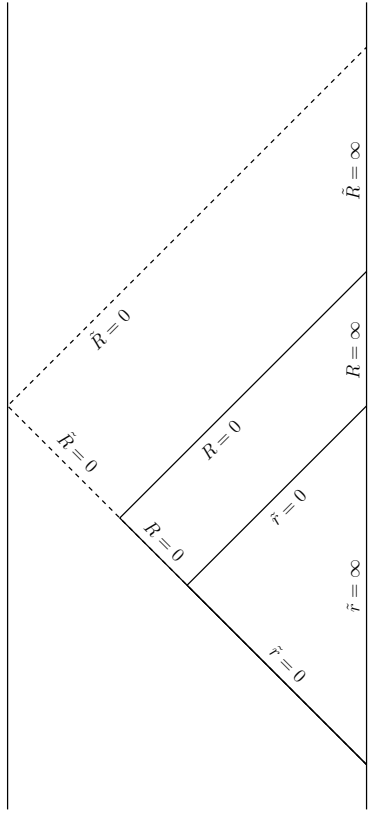

First embed a near-NHEK patch,
\be
ds^2 = 2M^2 \Gamma(\theta)\left[ -\tilde{r}(\tilde{r}+2\kappa) d\tilde{t}^2 + \frac{d\tilde{r}^2}{\tilde{r}(\tilde{r}+2\kappa)} + d\theta^2 + \Lambda(\theta)^2 (d\tilde{\phi} + (\tilde{r}+\kappa) d\tilde{t})^2 \right] \,,
\label{nearNHEK 3}
\ee
inside a NHEK patch \eqref{NHEK} as shown in Figure \ref{appfig1}. The transformation between the coordinates $(\tT,\tR,\tP)$ and $(\tilde{t},\tilde{r},\tilde{\phi})$ may be found in \cite{Hadar:2014dpa}. Following that up by a translation $\tilde{T}\to\tilde{T}-\chi$ one finds the following mapping between the near-NHEK patches \eqref{nearNHEK 1} and \eqref{nearNHEK 3}. 
\bea\label{NHEK time translation}
R &=& \tilde{r} + \chi e^{\k \tilde{t}} \sqrt{\tilde{r}(\tilde{r}+2\kappa)} \,, \non \\
T &=& \tilde{t}+\frac{1}{2\k} \ln \frac{\tilde{r}(\tilde{r}+2\kappa)}{R (R + 2\k)} \,, \\
\Phi &=& \tilde{\phi} - \frac{1}{2} \ln \frac{(R + 2\k)\tilde{r}}{R (\tilde{r}+2\k)} \,. \non
\eea
\noindent 
Note that the transformation of the circular orbit \eqref{circular orbit} in \eqref{nearNHEK 1} gives rise to orbits in \eqref{nearNHEK 3} which are qualitatively very similar to the ones studied in the paper. The only difference is that these orbits enter the throat in the infinite near-NHEK past ($\tilde{t}=-\infty$). In the terminology of \cite{Hadar:2015xpa} these are ``slow'' plunging or osculating orbits.

\begin{figure}[h!]
	\centering
	
	\resizebox{0.3\textwidth}{!}
	{
		
\begin{tikzpicture}
\draw[thick] (0,-5) -- (0, 9);
\draw[thick] (-8,-5) -- (-8, 9);

\draw[thick] (0,-4) to (-4, 0) to (0,4);

\draw[thick] (0,0) to (-4, 4) to (0,8);

\draw (-2,-1.5) node[rotate=-45]{$r=0$};
\draw (-2.75, 0.75) node[rotate=45]{$r=0$};
\draw (-0.25, -1.75) node[rotate=90]{$r=\infty$};

\draw (-2.75,3.25) node[rotate=-45]{$\tilde{r}=0$};
\draw (-2, 5.5) node[rotate=45]{$\tilde{r}=0$};
\draw (-0.25, 5.75) node[rotate=90]{$\tilde{r}=\infty$};

\end{tikzpicture}

	}
	
	\caption{Penrose diagram illustrating the setup for deriving the transformation \eqref{global time translation}. The wedge bounded by $\tilde{r}=0$ and $\tilde{r}=\infty$ is near-NHEK. The wedge bounded by $r=0$ and $r=\infty$ is also near-NHEK. The two near-NHEK wedges are related to each other by a global time translation $\tau\to\tau-\pi/2$.}
	\label{appfig2}
\end{figure}
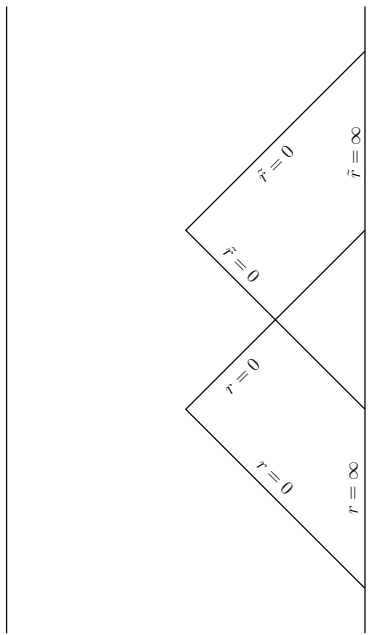

In this paper we considered the radiation from ``fast'' plunging or osculating orbits. These enter the throat at some finite near-NHEK time (which we set to $t=0$). Such ``fast'' orbits may be obtained by a transformation corresponding to a global time translation  $\tau\to\tau-\pi/2$, illustrated in Figure \ref{appfig2}. This mapping between the near-NHEK patches \eqref{nearNHEK 3} and \eqref{nearNHEK 2} is given by
\bea\label{global time translation}
\tilde{r} &=& \sqrt{r(r+2\kappa)}\sinh\kappa t-\k \,, \non \\
\tilde{t} &=& \frac{1}{2\k} \ln \frac{\sqrt{r(r+2\kappa)}\cosh\kappa t-(r+\k)}{\sqrt{r(r+2\kappa)}\cosh\kappa t+(r+\k)} \,, \\
\tilde{\phi} &=& \phi+ \half\ln\frac{(r+\k)\sinh\kappa t+\k\cosh\kappa t}{(r+\k)\sinh\kappa t-\k\cosh\kappa t}\,. \non
\eea
Note that the transformation \eqref{global time translation} corresponds to the $\chi=0$ case of \eqref{transformation}. 

The transformation \eqref{transformation} is a composition of \eqref{NHEK time translation} with \eqref{global time translation}.

\end{document}